# Research on Brightness Measurement of Intense Electron Beam[*]

WANG Yuan(王远), JIANG Xiao-Guo(江孝国), ZHANG Huang(张簧), YANG Guo-Jun(杨国君),
LiYi－Ding（李一丁）Li Jin(李劲))

Institute of Fluid Physics, China Academy of Engineering Physics, P.O.Box 919-106, Mianyang 621900,China

**Abstract** The mostly research fasten on high emission density of injector to study electron beam's brightness in LIA．Using the injector(2MeV) was built to research brightness of multi-pulsed high current(KA) electron beam, and researchs three measurement method(the epper-pot method、beam collimator without magnetic field、beam collimator with magnetic field method ) to detect beam's brightness with time-resolved measurement system.

**PACS**　78,29. 25.Dz, 29.27.Ac, 41.85.Ar　　　　**DOI**

## 1 Introduction

To study the emission characteristics of multi-pulse linear induction accelerators (LIA), it is important to measure the brightness of the emitted electron beams. Since the beam brightness of LIA is mainly determined by its injector, most of the beam brightness researches focus on the injector. In this paper, three methods of brightness measurement, combining with time-resolved measuring system, are benchmarked on a 2MeV 3kA LIA injector[1].

The time-resolved measuring system on LIA mainly measures the divergence, energy and transverse size of the intense electron beam. This optics diagnosing system is applied in the LIA to high time-resolved beam parameter measurement. It can obtain beam distribution at the exits of both the injector and the accelerator. It can assist the research on the correlation between magnetic field configuration and beam transport, the measurement of beam energy spectrum, et al. It can also provide abundant detailed image data to help the LIA commission[2−4].

## 2 Brightness measurement of intense electron beam

The method of beam brightness measurement changes for different beam distribution. For experimental measurement, only the uniform and the Gaussian beam distributions are convenient to measure. We studied three measurement methods: the epper-pot method, beam collimator without magnetic field method, beam collimator with magnetic field method.

The definition of beam brightness can be expressed as

$$B_n = \frac{\pi^2}{(\beta\gamma)^2} \frac{d^4 I}{dV_4} \quad (1)$$

which equals to $\pi^2$ times the density in four-dimension transverse phase space. In equation (1), $B_n$ is the normalized brightness, $d^4 I$ is the current element in four-dimension transverse phase space, $dv_4$ s the volume element of four-dimension transverse phase space, $\beta = \frac{V}{C}$ is the electron's relative velocity and $\gamma = (1-\beta^2)^{-\frac{1}{2}}$ is the relativistic factor. According to this definition, for a beam of fixed energy, the normalized brightness $B_n$ is determined by $d^4 I$ and $dv_4$. The beam brightness can be increased by raising $d^4 I$ and decreasing $dv_4$.

### 2.1 Beam emittance measurement

If the transverse distribution of electron beam is

*Supported by National Natural Science Foundation of China (10675104, 51077119, 11375162)
Email: ideawy@163.com.　　Tel: 0816-2484140

elliptic with the boundary equation as [5]

$$1-(x^2+y^2)/b^2-(x'^2+y'^2)/v^2=0 \quad (2)$$

the four dimension space volume is

$$V_4=\pi^2\varepsilon^2/2 \quad (3)$$

and the core emittance is

$$\varepsilon=bv \quad (4)$$

In practice the transverse beam distribution are almost uniform, so according to equation (1) the normalized brightness is

$$B_n=\frac{2I}{(\beta\lambda\varepsilon)^2}=\frac{2I}{\varepsilon_n^2} \quad (5)$$

where $\varepsilon_n$ is the normalized emittance. Once the current I and the normalized emittance $\varepsilon_n$ have been measured, the brightness Bn can be calculated by equation (5). The beam current can be measured by Faraday cup, small coupling coil or small resistance, and the emittance measurement is usually by multi-slits method or Perforated Plates method. In Perforated Plates measurement, a majority of the beam is absorbed by the Perforated Plates as depicted in figure 1, while the residual reaches a reflector and is detected by the time-resolved diagnosing system. From the image of the beam passing through, $\gamma$、$\theta$、$\delta_\theta$ can be obtained, and the phase ellipse can be drawn as shown in figure 2. Its area A divided by $\pi$ equals $B\cdot V$.

In practice, the electron distribution is Gaussian rather than uniform, and there is no clear-cut boundary in detected image. Because of the aberration, the detected gray image is also Gaussian and its boundary is regarded as the contour curve of 5% gray scale of the maximum. The beam radius can also be determined by the half maximum contour. The emittance obtained by this method is called the RMS emittance $\varepsilon_m$, which is about 1/3 of the core emittance theoretically:

$$\varepsilon=3\varepsilon_m \quad (6)$$

Thus the brightness is

$$B_n=\frac{2I}{9(\beta\gamma\varepsilon_m)^2} \quad (7)$$

The Perforated Plate thickness is about equal to the electron's penetrating range. To reduce the scattering, its material is usually graphite, aluminum or tantalum. When the beam energy is high, stainless steel is also used. The scintillate screen is plated on an aluminum foil of thickness ~0.1－0.3mm.

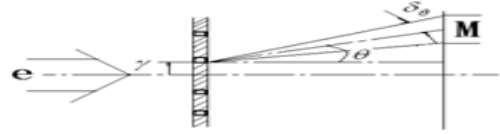

Fig.1 Emittance measurement by Perforated Plates

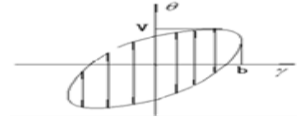

Fig.2 Phase ellipse

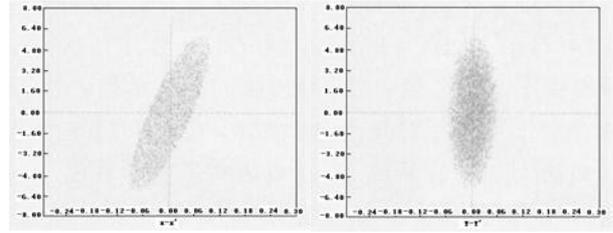

Fig.3 The X-X' and Y-Y' phase planes

The nonlinear effect is analyzed by Lie algebra method to second order in the case of K-V distribution. In computation, static electric lens field is divided into several small intervals which are treated as uniform accelerating field, and each dividing point is considered as a thin lens. The nonlinear electron trajectories can be obtained by applying the Lie maps of uniform accelerating field and thin lens in order. Figure 3 is the computed x - x' and y - y' phase planes.

**2.2 Brightness measurement without magnetic field**

The collimator without magnetic field is a straight pipe of length L and radius $R_p$, which is usually substituted with two circular collimator of distance L. For the orbit of electrons should be straight lines, the initial states (at the entrance of the collimator) of the electrons passing through should fulfill the condition

$$(X + X'L)^2 + (Y + Y'L)^2 \leq R_P^2 \quad (8)$$

Thus the four dimension volume is

$$V_4 = \pi^2 R^4 / L^2 \quad (9)$$

and the brightness can be expressed as

$$B_n = \frac{IL^2}{(\beta\gamma)^2 R_P^4} \quad (10)$$

according to equation (1). Once the current I of the exiting electron beam was measured, the beam brightness could be calculated by equation (10).

In practice, as a result of space charge effect and residual coil field, the electron trajectories in the collimator are helical curves rather than straight line. The straight trajectory condition is approximately fulfilled only when the collimator length L is negligible comparing to the convolution wavelength, which means

$$\int_0^L K_c dz \ll 2\pi \quad (11)$$

where $K_C$ is the convolution wave number

$$K_C = \frac{eB}{\beta\gamma mc^2} \quad (12)$$

The collimator method works only when the beam is dominated by emittance, which requires

$$B_n \ll \frac{I_a}{R_P^2} \quad (13)$$

where $I_a = \beta\lambda I_0$ is Alfvén current. There are also three other conditions to guarantee the collimator performance. Firstly, the beam radius $R_b$ should be larger than the pipe radius $R_p$ to be properly collimated:

$$R_b > R_P \quad (14)$$

For Secondly the thermal divergence of electron beams ($\varepsilon/R_b$) should be larger than the envelope angle ($R'$), namely $\varepsilon/R_b \gg R'$, in order to subdue the collimator's effects on the beam divergence. Thirdly, the maximum angle of electrons passing through the collimator ($2R_p/L$) must be smaller than the typical thermal angle ($\varepsilon/R_b$) in order to collimate the electron beam in the $X' - Y'$ space, thus

$$R_P < R_b < \frac{\varepsilon L}{2R_P} \quad (15)$$

and

$$R' \ll \frac{\varepsilon}{R_b} \quad (16)$$

should be fulfilled. For $B_n = \frac{2I}{\varepsilon_n^2}$, conditions (15) and (16) are equivalent to

$$R_P < R_b < \sqrt{\left(\frac{I}{2B_n}\right)} \frac{L}{\beta\gamma R_P} \quad (17)$$

And

$$R' < \sqrt{\left(\frac{2I}{B_n}\right)} / \beta\gamma R_b \quad (18)$$

**2.3 Brightness measurement with magnetic field collimator**

Another beam brightness measuring method is using a collimator pipe in uniform solenoid magnetic field. The electrons convolute in the pipe with the volume of trajectory spaces as $V_4 = \pi^2 K_C^2 R_P^4 / 6$. By measuring the passing through current I, the brightness can be obtained by

$$B_n = \frac{6I}{K_o^2 R_P^4} \quad (19)$$

where, $K_o = \frac{eB}{mc^2}$. Equation (19) doesn't including the beam energy.

The conditions (13), (16) and (18) should also be fulfilled, and the condition (15) is replaced by

$$R_P < R_b < \frac{\varepsilon}{K_c R_P} \quad (20)$$

for the angular acceptance of magnetic collimator is $K_c R_p$.

For magnetic collimator, the collimator thickness should be larger than the revolution wavelength to collimate all the electrons.

The above mentioned three brightness measuring methods are all applied on our experiments on LIA to benchmark each other.

## 3 Time-Resolved Measurement System

The time and space resolving power and imaging dynamic range are important parameters of the measurement system performance. ICCD is the key part of the camera, as is shown in Table 1. It is typically composed of a micro-channel plank (MCP), a coupling lens, and a CCD camera. As an image gating part, the high-speed MCP with a 25mm available cathode diameter and 2ns shutter time is utilized. The drift spaces located both upstream and downstream of the magnetic analyzer are 52mm long.

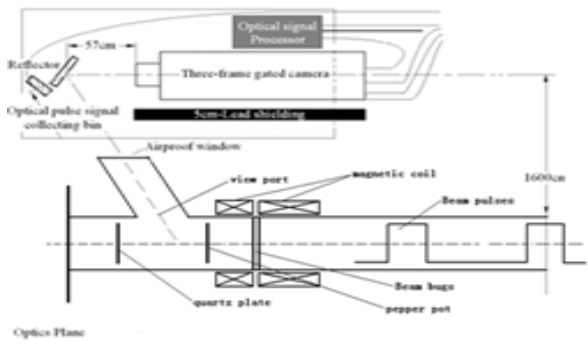

Fig. 4 Measurement based on magnetic analyzer method

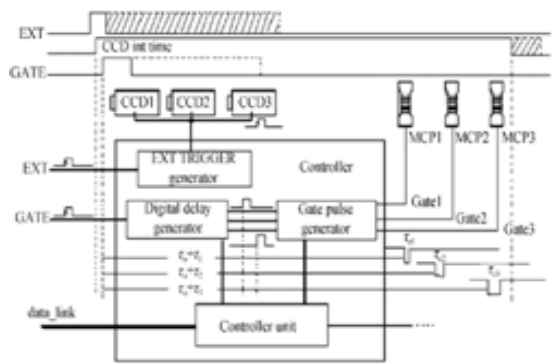

Fig. 5 Flow chart for the camera system control

There are two trigger inputs: one external trigger synchronization signal (EXT) and one image increasing gating trigger signal (GATE). The synchronization signal is obtained by a pulse acquisition device in the accelerator cell, which gets the current pulse (or prepulse) loaded from pulse power system and passes it to the trigger equipment. Delayed for a certain time, the signal is utilized as the input of multi-frame gated synchronization monitor model. After an inherence delay, the framing camera turns on its gate and begins to photograph. With a proper delayed time, the image of pulse electron beam will just fall into the gated duration of the camera[6-10].

Table 1 Main Performance of Three-Frame Gated ICCD

|  | Min | Max | tep |
|---|---|---|---|
| Exposure Time(ns) | 2 | ~109 | <0.5 |
| Interval Time(ns) | ~0 | ~109 | <0.5 |
| Spatial Resolution(p.mm) |  | >30 |  |
| Equivalent Background Lamination(EBI) (electrons pixel s) |  | 10 |  |
| Response Uniformity |  | 2% |  |
| Linearity |  | 1% |  |
| Inherent Insertion Delay(ns) | 36.8 | 37.3 |  |
| Image Size In Diameter(mm) |  | 25 |  |

## 4  Experiment Result

A Perforated Plate of cross places pinholes was using in the experiments. It's a tantalum plate of thickness 3mm and effective diameter 80mm. There were nine pinholes horizontally and vertically (as show in figure 6). The luminous body was quartz glass, which provided electron beam information by Cerenkov radiation. Its optical information was recorded by time-resolved measurement system. Since the duration of Cerenkov radiation is only of sub-nanosecond level, the measurement system could distinguish different beam pulses.

In our experiment, the multi-pulsed beam energies were 2MeV (single pulsed), 1MeV (double pulsed), and 500KeV (triple pulsed). For quartz glass (n＝1.5), these beam energies are enough to produce Cerenkov radiation. The angle is (2MeV), (500KeV), and (500KeV) correspondingly. The experimental layout is shown in figure 1. Figure 6 is the integral images of the multi-pulsed experiment recorded by the time-resolved measurement system, with each pulse duration of ~100ns and pulse spacing around 300ns.

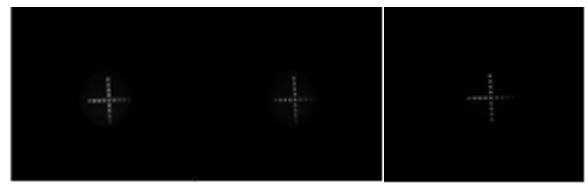

Fig. 6 photo of integral spots of time-resolved measurement, time interval 10ns

The data processing is based on the single-pulse integral emittance data analysis. A majority of electron beam is absorbed by the Perforated Plates, and the rest, which pass through the pinholes, reaches a quartz screen and produces Cerenkov radiation (as depicted in figure 1). The time-resolved optical diagnosing system records the radiation and obtains the RMS radius and divergence of the beam. The beam emittance can also be obtained by analyzing the data recorded at some given time. According to the analysis, the electron beam's normalized brightness is $4.14 \times 10^8 A/(m.rad)^2$.

# 高密度强流束亮度测量技术研究

王远，江孝国，张篁，杨国君，李一丁，李劲

（中国工程物理研究院 流体物理研究所 106 室，四川 绵阳 621900）

**摘要** 直线感应加速器产生强流高亮度束的研究主要集中于强流高亮度注入器的研究。利用已经建成的 2MeV LIA 注入器，通过研究多脉冲产生强流（KA）高亮度 电子束，并结合时间分辨测量系统研究了三种测量技术：发射度测量法、无场准直器和磁场准直器的测量方法。

**关键词** 直线感应加速器（LIA）;时间分辨测量; 高亮度电子束; 测量技术

**中图分类号** TL53 **文献标识码：A**